

\input phyzzx

%
\titlepage
\title{FEYNMAN RULES IN THE TYPE III $\rm NATURAL^*$ FLAVOUR-CONSERVING
TWO-HIGGS DOUBLET MODEL}
\author {Chilong Lin, Chien-er Lee, and Yeou-Wei Yang}
\address{Department of Physics, National Cheng Kung University, Tainan,Taiwan,
Republic of China}

\abstract
  We consider a two Higgs-doublet model with
$S_3$ symmetry, which implies a $\pi \over 2$ rather than 0 relative phase
between the vacuum expectation values
$<\Phi_1>$ and $<\Phi_2>$.
The corresponding Feynman rules are derived accordingly and
the transformation of the Higgs fields
from the weak to the mass eigenstates includes not only an angle rotation
but also a phase transformation. In this model,
both doublets couple to the same type of fermions and the flavour-changing
neutral currents are naturally suppressed. We also demonstrate that the Type
III natural flavour-conserving model is valid at tree-level even when an
explicit $S_3$ symmetry breaking perturbation is introduced to
get a reasonable CKM matrix.
In the special case $\beta = \alpha$, as
the ratio $\tan\beta = {v_2 \over v_1}$ runs from 0 to $\infty$, the
dominant Yukawa coupling will change from the first two generations
to the third generation. In the Feynman rules, we also find that the
charged Higgs currents are explicitly left-right asymmetric. The
ratios between the left- and
right-handed currents for the quarks
in the same generations are estimated.

* This work was supported by the National Science Council of the Republic of
Chi

\endpage

\chapter{INTRODUCTION}

  The two-Higgs doublet $\rm model^{1-8,13,14}$ is a rather popular extension
of the
standard model. Introducing another doublet would lead to the flavor-changing
neutral $\rm current^{4-8,11}$ (FCNC) problem in its Yukawa coupling sector.
There are two ways to avoid this problem, $\rm which^{11}$ are generally called
the two types of natural flavour conserving (NFC) models. $\rm One^{4-6,11}$
is to let only one Higgs doublet couple to the fermions and the other
to let the doublets couple seperately to different types of fermions.
In our previous $\rm
investigations^{7-10}$, we found another way in which both doublets can
couple to all types of fermions without FCNC
by introducing $S_3$ permutation symmetry. In that model,
both doublets couple to all types of fermions and the relative phase
$\theta$ between the
vacuum expectation values (VEV's) of the doublets is found to be $\pi \over 2$.
The coupling constant matrices corresponding to the doublets can be
diagonalized simultaneously by the same transformation so that the FCNC problem
was naturally solved in the $S_3$ model.

 In the two-Higgs doublet models, the relative phase $\theta$ between the
VEV's could be either 0 or $\pi \over 2$.
The $\theta$=0 case has been considered by $\rm Bertolini^5$. Since
the $S_3$ model implies $\theta={\pi \over 2}$, our main interest
here is to investigate this case,
where we have a pure imaginary vacuum expectation value (VEV) for
the second Higgs doublet $\Phi_2$.
The transformation from the weak to the mass eigenstates of the Higgs fields
will be altered considerably by the imaginary property of $<\Phi_2>$.
The detail is given in section 2.

  In most two-Higgs doublet models, the Yukawa couplings of the Higgs
doublets to the fermions can not be diagonalized simultaneously, which
lead to the FCNC problem. We present another natural flavour-conserving (NFC)
model with explicit $S_3$ symmetry breaking in section 3.
The CKM matrix is also discussed there.

  Since $<\Phi_2>$ becomes imaginary in the $S_3$ model,
the Feynman rules in
ref. [5] should be modified. The derivation is given in section 4. Since
one can couple both doublets to the same type of fermions without FCNC,
the derived Feynman rules contain contributions from both doublets and
are very different from those given before. We also find that
the dominant contribution to the Feynman rules will change from the first
(lighter) two generations to the third (heavier) generation of fermions
as the ratio ${v_2 \over v_1}=\tan\beta$ varies
from 0 to $\infty$. We conclude this paper in section 5.

\chapter{THE $\theta={\pi \over 2}$ CASE IN THE TWO-HIGGS DOUBLET MODEL}

  The mass spectra and physical eigenstates of Higgs fields have been
illustrated in ref. [5] on a simplified two-Higgs doublet model
which takes the relative phase $\theta=0$.
In general, one needs to introduce an additional symmetry under which the
doublets transform differently so as to distinguish the doublets.
The most popular transformation is
$\Phi_1 \to \Phi_1$, $\Phi_2 \to -\Phi_2$ and the corresponding
general $SU(2)_L \times U(1)_Y$ gauge-invariant, renormalizable
Higgs potential is given by $Bertolini^5$ and $Toussaint^{12}$.

  In breaking the original $SU(2)_L \times U(1)_Y$ gauge symmetry down to
$U(1)_{EM}$, the VEV's are chosen as
$<\Phi_1>={v_1 \over \sqrt2}$ and $<\Phi_2>={{v_2 e^{i\theta}}\over \sqrt2}$,
where $\theta$ is the phase difference between the vacuum expectation values.
After SSB, the Higgs fields are written as
$$\Phi_1  = \left(\matrix{{H_{1c}\over \sqrt2} \cr
            {1\over{\sqrt2}}(v_1 + R_1 +iI_1)}\right),
\Phi_2  = \left(\matrix{{H_{2c} \over \sqrt2} \cr
            {1\over{\sqrt2}}(iv_2+ R_2
          +iI_2)}\right).\eqno(2.1)$$
where $H_{ic}$'s are the charged Higgs fields and $R_i$'s ($I_i$'s) are the
real (imaginary) parts of the neutral Higgs fields.
The minimization condition ${\partial V} \over {\partial \theta}$=0 tells us
that $\theta$ can only be 0 or $\pi \over 2$. The case $\theta$=0 was
discussed in ref. [5]. In this paper, we shall concentrate on the
case $\theta~= ~{\pi \over 2}$, which change the sign of $\lambda_5$ in
ref. [5] and mix the real and imaginary parts of the neutral Higgs fields. Here
is the summary:

 (a). The charged components of $\Phi_1$ and $\Phi_2$ mix to give a charged
Goldstone boson $G^+$ and a physical charged Higgs $H^+$.
The transformation U from their weak eigenstates to mass eigenstates
includes not only an angle rotation but also a phase transformation as follows
$$   U  = \left(\matrix{cos{\beta} & sin{\beta} \cr -sin{\beta} &
      cos{\beta} \cr} \right) \left(\matrix{1 & 0 \cr 0 & -i \cr} \right)
                                         \eqno(2.2) $$
where $\tan\beta = {v_2 \over v_1}$.

 (b). For $\theta={\pi \over 2}$, the constructions of the neutral
Higgs fields are different to that for $\theta
=0$. In the $\theta=0$ case, $R_1$ and $R_2$ mix to give the scalar
fields and $I_1$, $I_2$ mix to give the pseudoscalar ones.
But, in the $\theta={\pi \over2}$ case, $R_1$ and $I_2$ are combined
to give the scalar fields while $I_1$ and $R_2$ are combined to give the
pseudoscalars. The real part and the imaginary part mix together.
Therefore, the trasformation between the weak and mass eigenstates
also need an additional phase transformation like that in (2.2).

  For pseudoscalar fields, $I_1$ and $-iR_2$ mix to
give a Goldstone boson $G^0$ and a physical pseudoscalar $H_3^0$ with the
same U as in the charged Higgs fields. For the scalar fields, $R_1$ and $iI_2$
mix to give the scalar fields $H_1^0~ and~H_2^0$ with the transformation
$$ \left(\matrix{H_1^0 \cr H_2^0 }\right)
       = \left(\matrix{cos{\alpha} & sin{\alpha} \cr -sin{\alpha} &
          cos{\alpha} \cr} \right)
         \left(\matrix{1 & 0 \cr 0 & -i \cr} \right)
         \left(\matrix{R_1 \cr iI_2 }\right) \eqno(2.3) $$
where $\tan{\alpha}= {{-(A-B)+ \sqrt{(A-B)^2+C^2}} \over C}$ with A, B and C
defined in ref. [5].

  In the former diagonalization of the Higgs fields, it is found that the
charged Higgs and the pseudoscalars have the same mixing angle.
But the mixing angle for the neutral scalars is different.
One may define a set of new doublets $\Phi_1'$ and $\Phi_2'$ in which the
charged Higgs fields and the pseudoscalars are in their mass eigenstates
while the scalars $\phi_1^0$, $\phi_2^0$ are not.
$$\Phi_1'= \Phi_1 cos{\beta} -i \Phi_2 sin{\beta}
  = \left(\matrix{G^+ \cr v+{1\over{\sqrt2}}(\phi_1^0 +iG^0)}\right ),$$
$$\Phi_2'=-\Phi_1 sin{\beta} -i \Phi_2 cos{\beta}
  = \left(\matrix{H^+ \cr {1\over{\sqrt2}}(\phi_2^0
   +i H_3^0)}\right).\eqno(2.4)$$
where $v={1 \over 2}(v_1^2 +v_2^2)$. The relation between
$H_1^0$, $H_2^0$ and $\phi_1^0$, $\phi_2^0$ is
$$ \left(\matrix{H_1^0 \cr H_2^0 }\right)
       = \left(\matrix{cos(\alpha-\beta) & sin(\alpha-\beta) \cr
         -sin(\alpha-\beta) & cos(\alpha-\beta) \cr} \right)
         \left(\matrix{\phi_1^0 \cr \phi_2^0 }\right). \eqno(2.5) $$

  If $\beta = \alpha$, then all the Higgs fields are diagonalized
simultaneously by the transformation (2.2).

\chapter{THE $S_3$ MODEL AND ITS FCNC}

  In standard model, there is only one
Higgs doublet whose phase of VEV can be rotated away by a gauge transformation.
Therefore, no spontaneous CP-violation will appear in standard model. It was
widely suggested that one needs at least one more Higgs doublet to
produce
CP-violation spontaneously. But, most two-doublet models meet the FCNC
problem which arises from the non-simultaneous diagonalization of the coupling
constant matrices those couple to different doublets respectively. Glashow
and $\rm Weinberg^{11}$ suggested two ways to $avoid$ the FCNC problem.
One is to let the doublets couple seperately
to different types of fermions. The other is to let only one
doublet couples to both types of fermions. In our previous $\rm
investigations^{7-10}$, we found another way to suppress the FCNC naturally
with an additional $S_3$ permutation symmetry.

  The motivation of introducing the $S_3$ symmetry is based on the similarity
between the three generations of fermions. For the same type of fermions,
they are very similar to each other except their masses. It is reasonable to
assume that there is no fermion masses before SSB and thus no difference
between the generations. For three generations of fermions, $S_3$ symmetry is
conserved before SSB. When SSB happen, the fermions get their masses and the
$S_3$ symmetry was $spontaneously$ broken. In our $S_3$ model, the
fermions are classified to the three-dimensional
$\rm representations^{8}$
$\Gamma^6$ of $S_3$, which is just the change of generations.
The Yukawa Lagrangian with both doublets coupled to the down-type right-handed
fermions is then written as
$$\eqalign{L_{YUK}^d =& \,
                \bar{Q'_{L}} (\Phi_1 G_1 + \Phi_2 G_2) D'_R +H.C. \cr
         = & \, (\bar{U} V, \bar{D})_L [\Phi'_1 (c G'_1 +i s G'_2)+
                  \Phi'_2(-sG'_1 +icG'_2)] D_R +H.C.\cr} \eqno(3.1)$$
where $Q'_L=(U',D')_L$ means the left-handed quark fields in the weak
eigenstates and V is the CKM-matrix.
The unprimed $Q_L,~U_L~and~D_L$ are in their mass eigenstates.
The unprimed $G_1,~G_2$ are the Yukawa coupling constant matrices for the down
type quarks
and couple to $\Phi_1$ and $\Phi_2$, respectively. The primed coupling
constant
matrices are defined as $G'_i=U_d G_i U_d^{\dagger}$ that are diagonal. The
primed Higgs doublets is defined in (2.4) and $c=\cos\beta$,
$s=\sin\beta$ for simplicity. The mass matrix of the
down type quarks under the spontaneous $S_3$ breaking model is exppressed
as $$\eqalign{ M_d= & \, M_{d1} + M_{d2} = <\Phi_1>G_1 +<\Phi_2>G_2 \cr
        = & \,    <\Phi_1> \left(\matrix{a & b & b \cr b & a & b
      \cr b & b & a}\right)
 +<\Phi_2> \left(\matrix{0 & -d & d \cr d & 0 & -d \cr -d & d & 0 \cr}
 \right), \cr
   = & \,\left(\matrix{A & B-iD & B+iD \cr B+iD & A & B-iD \cr B-iD &
        B+iD & A}\right),}\eqno(3.2)$$
where $A={av_1 \over \sqrt{2}}$, $B={bv_1 \over \sqrt{2}}$ and $D={dv_2 \over
\sqrt{2}}$ and the phase of $\Phi_2$ must be $\pi/2$. Since $G_1$ and $G_2$ can
be diagonalized simultaneously by the same trasformation matrix. The FCNC
problem does not appear at tree-level.

  The matrix form of the up type quarks are similar to (3.2) with A, B and D
replaced by $A'$, $B'$ and $D'$.
The eigenvalues (quark masses) are given as follows
$$\eqalign{ & \, m_d=A-B-\sqrt{3}D,~~~~m_s=A-B+\sqrt{3}D,~~~~m_b=A+2B \cr & \,
  m_u=A'-B'-\sqrt{3}D',~~~~ m_c=A'-B'+\sqrt{3}D',~~~~m_t=A'+2B'} \eqno(3.3)$$

 Substituting the experimental current quark masses into these
expressions and taking the unknown top quark mass to be 135 GeV, we obtain the
values of the parameters as follows
$$\eqalign{& \, A=1.828 GeV,~~~~ B=1.736 GeV,~~~~ D=0.048 GeV, \cr & \,
  A'=45.45 GeV,~~~~ B'=44.77 GeV,~~~~ D'=0.388 GeV} \eqno(3.4)$$
where we have used the quark masses data: $m_u=0.0051$,
$m_d=0.0089$, $m_s=0.175$, $m_c=1.35$ and $m_b=5.2$ in GeV.

  After SSB, the Yukawa Lagrangian for the down type right-handed
quarks is given by
$$\eqalign{L_{Yuk}^d & \,
         = (\bar{U} V,\bar{D})_L  \cr
           & \, \times \left(\matrix{{gM^d_d
                \over {\sqrt{2}M_W}} G^+ +{g\over {\sqrt{2}M_W}}(-M^d_{d1}
               \tan{\beta} +M^d_{d2} \cot{\beta}) H^+ \cr
               M^d_d + {{g M^d_d} \over {2M_W}}(\phi_1^0 +iG^0)
              +{g \over {2M_W}}(-M^d_{d1} \tan{\beta} +M^d_{d2} \cot{\beta})
               (\phi_2^0 + iH_3^0)} \right) \cr
           & \, \times D_R+H.C.} \eqno(3.5)$$
where the superscript d on the mass matrices means diagonalized .
The Yukawa Lagrangian for the up type quarks can be derived in a similar way.

 In the above discussions, the mass matrices corresponding to different Higgs
doublets are diagonalized by the same unitary transformation and thus no FCNC
problem at all. But it also leads to a $3 \times 3$ unit CKM matrix.
In ref. [8], we added an explicit $S_3$
breaking P' to the up type quarks as a perturbation of the originally
spontaneously broken $M_u^{(0)}$ to get a reasonable CKM matrix.
Since we did not add any perturbation to $M_d^{(0)}$, the neutral currents of
down type quarks are still flavour-conserving at tree level.
In what follows, we shall demonstrate that the neutral currents of up type
quarks is also NFC at tree level if the perturbation is chosen suitably.

 We devide the perturbation $P'$ into $P'_1$ and $P'_2$ which correspond
to the doublets respectively. The mass matrices are defined as follows
$$M_u=M_u^0+P'=M^0_{u1}+P'_1+M^0_{u2}+P'_2,~~~~~~~M_d=M^0_d=M^0_{d1}
      +M^0_{d2}     \eqno(3.6) $$
where $M^0$ are the spontaneous $S_3$ breaking matrices and $P'_i$'s are the
perturbations. The matrix $M_u$ is diagonalized in two stages. In the first
stage, $M^0_{d1}$, $M^0_{d2}$, $M^0_{u1}$ and $M^0_{u2}$ are simultaneously
diagonalized by $U^{(0)}$ and $P'_i$ transform to $P_i$.
$$U^{(0)} M_u U^{(0)\dagger} =D^0_u+P=D^0_{u1}+P_1+D^0_{u2}+P_2$$
$$U^{(0)} M_d U^{(0)\dagger} =D^0_d=M^d_{d1}+M^d_{d2}=Diad.(m_d,m_s,m_b)
    \eqno(3.7)$$
The matrix $D^0_u+P$ is then diagonalized by $U^{(n \ge 1)}=V$ in the
second stage
$$M_u^{diag}= V (D_u^0+P) V^{\dagger}=(M'_{u1}+M^P_{u1})+(M'_{u2}
    +M^P_{u2})=Diag.(m_u,m_c,m_t) \eqno(3.8)$$
where $U^{(n)}$ is the correction up to the n'th order in the perturbation.

The Yukawa couplings of $H_1^0$ and $G^0$ are proportional to $M_u^{diag}$ so
that the corresponding neutral currents are NFC
while the couplings of $H_2^0$ and $H_3^0$ are
proportional to
$$-{\sin{\alpha} \over \cos{\beta}}(M'_{u1}+M^P_{u1})+
        {\cos{\alpha} \over \sin{\beta}}(M'_{u2}
    +M^P_{u2}) \eqno(3.9)$$
which should be diagonal for the sake of NFC. Since both (3.8) and
(3.9) are diagonal, $(M'_{u1}+M^P_{u1})$ and $(M'_{u2}+M^P_{u2})$ must
also be both diagonal. Since $D^0_{u1}$ and $D^0_{u2}$ are known and
$U^{(n)}=V$ is calculated in the previous $\rm investigation^8$, we can
calculate
$M'_{u1}$ and $M'_{u2}$ whose off-diagonal elements are just
canceled by those of $M^P_{u1}$ and $M^P_{u2}$ respectively.
Therefore, the simplest choice of $M^P_{u1}$ and $M^P_{u2}$ is
$$ M_{u1}^P = \left(\matrix{0 & 0 & -E^* 3B' \cr 0 & 0 &
-F^* 3B' \cr -E 3 B' & -F 3 B' & 0} \right) $$
$$ M_{u2}^p = \left(\matrix{0 & -2D^* \sqrt{3}D' & -E^* \sqrt{3}D' \cr
-2D \sqrt{3}D' & 0 & F^* \sqrt{3}D' \cr -E \sqrt{3}D' & F \sqrt{3}D'
& 0} \right) \eqno(3.10)$$
where, for simplicity, we use the first order result $U^{(1)}=V$.
The inverse transformations $V^{\dagger} M_{ui}^P V$ then give the
required
perturbations $P_i$ which give NFC neutral currents at tree level.
Thus, we can always choose explicit $S_3$ symmetry breaking perturbation which
preserves NFC at tree level and at the same time produces reasonable CKM
matrix.

 \chapter{THE MODIFIED FEYNMAN RULES}

  In this section, we derive the Feynman rules corresponding to the phase
difference $\theta={\pi \over 2}$ between the vacuum expectation values of the
Higgs fields. We find that the vertices involving the Higgs
boson-gauge boson trilinear interactions and the Higgs boson-gauge boson
four-point interactions are not modified. Only vertices
involving the fermion Yukawa coupling are modified.
The relevant modified Feynman rules of the quark Yukawa couplings in the
't Hooft-Feynman gauge are

$$\eqalign{ & \, {-ig \over {2M_W}}(M^d_{d2}
{sin{\alpha}\over sin{\beta}}+M^d_{d1} {cos{\alpha}\over cos{\beta}})
{}~~~~~~~~~~~~~~~~~~~~~~~~~~~~~~~~~~~~~~~~~~~~~ \cr
       & \, {-ig \over {2M_W}}(M^d_{d2} {cos{\alpha}\over
          sin{\beta}}-M^d_{d1} {sin{\alpha}\over cos{\beta}})   \cr
       & \, {g\gamma_5 \over {2M_W}}(M^d_{d2} \cot{\beta}-M^d_{d1}
          \tan{\beta})  \cr
       & \, {g\gamma_5 \over {2M_W}}M_D^d \cr
       & \, {-ig \over {2M_W}}(M^d_{u2}
       {sin{\alpha}\over sin{\beta}}+M^d_{u1} {cos{\alpha}\over cos{\beta}})
\cr        & \, {-ig \over {2M_W}}(M^d_{u2} {cos{\alpha}\over
          sin{\beta}}-M^d_{u1} {sin{\alpha}\over cos{\beta}}) \cr
       & \, {g\gamma_5 \over {2M_W}}(M^d_{u2} \cot{\beta}-M^d_{u1}
          \tan{\beta}) \cr
       & \, {-g\gamma_5 \over {2M_W}}M_U^d \cr
       & \, {{-ig V_{ij}}\over {2\sqrt{2}M_W}}[(1-\gamma_5)(M^d_{u1}
          \tan{\beta}-M^d_{u2} \cot{\beta})_{ii}+(1+\gamma_5)(M^d_{d2}
          \cot{\beta}-M^d_{d1} \tan{\beta})_{jj}]  \cr
         \cr
         \cr
       & \, {{-ig V_{ji}^{-1}\over {2\sqrt{2}M_W}}[(1+\gamma_5)(M^d_{u1}
          \tan{\beta}-M^d_{u2} \cot{\beta})_{ii}+(1-\gamma_5)(M^d_{d2}
          \cot{\beta}-M^d_{d1} \tan{\beta})_{jj}] }}$$

  We note that the quark Yukawa vertices corresponding to the
Goldstone bosons are the same as those in ref. [5],
but the vertices corresponding
to the physical Higgs bosons are considerably different. We also note
that the vertices of the charged Higgs fields depend only on $\beta$.

    In the case of $\beta=\alpha$, all Higgs fields are diagonalized
simultaneously. This leads to the following interesting resluts:
The $H_1^0$ vertices reduce to those given in ref. [5], while
those vertices corresponding to $H_2^0$, $H_3^0$ and $H^{\pm}$ become

$$\eqalign{ & \, {-ig \over {2M_W}}(M^d_{d2} \cot{\beta}-M^d_{d1}
               \tan{\beta})_{jj}~~~~~~~~~~~~~~~~~~~~~~~~~~~~~~~~~~~~\cr
       & \, {g\gamma_5 \over {2M_W}}(M^d_{d2} \cot{\beta}-M^d_{d1}
               \tan{\beta})_{jj} \cr
       & \, {-ig \over {2M_W}}(M^d_{u2} \cot{\beta}-M^d_{u1}
               \tan{\beta})_{ii} \cr
       & \, {-g\gamma_5 \over {2M_W}}(M^d_{u2} \cot{\beta}-M^d_{u1}
               \tan{\beta})_{ii} \cr}$$

   We observe that in the above vertices, all terms depend on
$X_i =(M^d_{u1}\tan{\beta}-M^d_{u2}\cot{\beta})_{ii}$ or
$Y_j =(M^d_{d1}\tan{\beta}-M^d_{d2}\cot{\beta})_{jj}$. The $X_i$ terms
always appear in the vertices involving the up-type quarks, while the $Y_j$
terms always appear in those involving the down-type quarks.

  In the natural flavour-conserving $S_3$ model, since $m_u$ and $m_d$ are too
small compared with other quarks, we may assume them to be zero, which lead to
$A-B=\sqrt{3} D=x$ and
$A'-B'=\sqrt{3} D'=x'$. Then we may express $X$ and $Y$ as follows

$$X =M^d_{u1}\tan{\beta}-M^d_{u2}\cot{\beta}=
   \left(\matrix{x'(\tan{\beta}+\cot{\beta}) & 0 & 0 \cr 0 &
 x'(\tan{\beta}-\cot{\beta}) & 0 \cr 0 & 0 & z'\tan{\beta} \cr} \right),
        $$
$$Y =M^d_{d1}\tan{\beta}-M^d_{d2}\cot{\beta}=
   \left(\matrix{x(\tan{\beta}+\cot{\beta}) & 0 & 0 \cr 0 &
   x(\tan{\beta}-\cot{\beta}) & 0 \cr 0 & 0 & z\tan{\beta} \cr} \right).
      \eqno(4.1)$$
where $z=A+2B$ and $z'=A'+2B'$.

  For the neutral $H_2^0$ and $H_3^0$ vertices, the relative magnitudes of the
vertices depend only on those of the diagonal elements of $X$ and $Y$.
One may find in (4.1) that $\tan{\beta}$
dominates the variations of the elements. We discuss only the following
three special cases here and the details are shown in Fig. (1) to (4).

  (1). $\tan\beta \to 0$:
  The couplings for the top and bottom quarks are very small.
The contributions only come from the first two generations. This
contradicts the general assumption of top and bottom dominance.
We also note that the vertex factors for the first two generations are
of different signs.

  (2). $\tan\beta \to \infty$:
  The top and bottom vertices dominate over the lighter ones by
about ${z'\over x'}\sim 200$ times for the up type and ${z\over x} \sim 60$
times for the down type, so we may neglect the first two generations.
This agrees the general assumption of heavy quark dominance.

  (3). $\tan\beta =O(10^{-1})$:
  Assuming that the couplings of the first two generations are of the
same order of magnitude as those of the third generation, then
$\tan^2{\beta} \sim {x\over z}~({x' \over z'})$ for the down (up) type
quarks. When $\tan\beta$ is of the order $10^{-1}$, no vertex
should be neglected.

     There are also something interesting in the charged Higgs mediated
currents.
Since $X_i$ and $Y_j$ varies as $\beta$ changes from 0 to $\pi \over 2$, the
relative magnitudes of the left-
and right-handed charged currents may also change.
For i=j, the ratio of the left-right currents in the $H^+$
vertices is about 7 to 1 for the first two generations and about 27 to 1
for the third generation. These ratios would be reversed in the $H^-$
vertices.
When $i \ne j$, the ratios of the left- and right-handed currents depend
on $\tan{\beta}$ and can not be determined.

\chapter{CONCLUSIONS AND DISCUSSIONS}

  In this paper, we present and discuss the Feynman rules for
the relative phase $\theta={\pi \over 2}$
instead of the generally considered $\theta=0$.
The roles of the Higgs fields are very different from those for
$\theta=0$. This non-vanishing $\theta$ leads to a pure imaginary VEV for
$\Phi_2$ and the real and
imaginary parts of the neutral Higgs mix to give the mass
eigenstates, i.e, $(R_1,I_2) \to (H_1^0, H_2^0)$
and $(I_1,R_2) \to (G^0, H_3^0)$. One needs additional phase transformation
to diagonalize the Higgs fields, which is absent in the $\theta=0$ case.
We also present a Type III NFC model with $S_3$ symmetry which is valid at
tree-level even when an explicit $S_3$ breaking perturbation is introduced
to get a reasonable CKM matrix.

   In the special case $\beta=\alpha$, we find that the generally-considered
heavy-quark
dominance is valid only when $\tan\beta$ is larger than 0.075 (0.13)
for the up (down) type quarks. When $\tan\beta$ is smaller than these
values, the first two generations dominate over the third genertaion.
The variations of $X_i$ ($Y_j$) in the Yukawa couplings of the up (down) type
quarks are shown in Fig. (1) and (2) (Fig. (3) and (4)) as functions of
$\beta$. We also noticed the left-right asymmetry in
the charged currents mediated by $H^{\pm}$. Independent of the ratio
$\tan{\beta}$
between $v_2$ and $v_1$, the i=j vertices are clearly left-right asymmetric.
The ratios of the left-right asymmetry in the $H^+$ mediated charged currents
are estimated to be about 7 to 1 for the first two generations and about 27
to 1 for the third generation. The ratios in the $H^-$ mediated currents
are reverse to those mentioned above. For $i \ne j$, the currents cannot be
adjusted to be left-right
symmetric simultaneously, and so there must be left-right asymmetry in the
charged Higgs mediated currents at tree level.

\endpage

\ref{T.D. Lee, Phys. Rev. D8, 1226(1973); Phys. Rep. 9, 143(1974).}
\ref{H.E. Haber, G.L. Kane, and T. Starling, Nucl.Phys. B161, 493(1979).}
\ref{K.S. Babu, and Ernest Ma, Phys. Rev. D31 2861(1985).}
\ref{S. Bertolini and A. Sirlin, Nucl. Phys. B248, 589(1984).}
\ref{S. Bertolini, Nucl. Phys. B272, 77(1986).}
\ref{G.C. Branco, A.J. Buras and J.M. Gerard, Nucl. Phys. B259, 306(1985).}
\ref{Chilong Lin, Chien-er Lee, and Yeou-Wei Yang, Chin. J. Phys. 26,
180(1988).}
\ref{Chien-er Lee, Chilong Lin and Yeou-Wei Yang Phys. Rev. D42, 2355(1990).}
\ref{Chilong Lin, Chien-er Lee and Yeou-Wei Yang, NCKU-HEP, Feburary,
   (1993).}
\ref{Chien-er Lee, Yeou-Wei Yang, Y-L. Chang, and S-N. Lai, Chin. J. Phys. 24,
   254(1986).}
\ref{S. L. Glashow and S. Weinberg, Phys. Rev. D15, 1958(1977).}
\ref{D. Toussaint, Phys. Rev. D18(1978), 1626.}
\ref{J. F. Gunion, H. E. Haber, G. Kane and S. Dawson,"The Higgs Hunter's
           Guide", (Addison-Wisley, 1990)}
\ref{M. Sher, Phys. Rep. 179(1989)273.}

\refout
\endpage

$$\rm       Figure~Captions  $$

Fig. (1). The $X_i$ factor in the Yukawa couplings of the up-type quarks with
        $H_2^0$ and $H_3^0$ as $\beta$ varies from 0 to $\pi \over 2$. The
        lines of the u and c quarks are so close that they overlap.

Fig. (2). The enlarged part of the intersecting region in Fig. (1). The light
        quarks dominate for $\beta <$ 0.075.

Fig. (3). The $Y_j$ factor in the Yukawa couplings of the up-type quarks with
        $H_2^0$ and $H_3^0$ as $\beta$ varies from 0 to $\pi \over 2$. The
        lines of the d and s quarks are so close that they overlap.

Fig. (4). The enlarged part of the intersecting region in Fig. (3). The light
        quarks dominate for $\beta <$ 0.13.

\end